*Application Note*

# SeqManager: A Web-Based Tool for Efficient Sequencing Data Storage Management and Duplicate Detection

Margot Celerie[1], Andrew Oldfield [1], William Ritchie [1,*]

[1] IGH, Univ Montpellier, CNRS, Montpellier, France

*Corresponding author. E-mail:William.ritchie@igh.cnrs.fr

**Abstract**
**Motivation:** Modern genomics laboratories generate massive volumes of sequencing data, often resulting in significant storage costs. Genomics storage consists of duplicate files, temporary processing files, and redundant intermediate data.
**Results:** We developed SeqManager, a web-based application that provides automated identification, classification, and management of sequencing data files with intelligent duplicate detection. It also detects intermediate sequencing files that can safely be removed. Evaluation across four genomics laboratory settings demonstrate that our tool is fast and has a very low memory footprint.
**Availability and implementation:** SeqManager is freely available under the MIT license at https://github.com/AIGeneRegulation/Sequencing-Data-Manager.

## 1 Introduction

Next-generation sequencing (NGS) technologies enable comprehensive analysis of genomes, transcriptomes, and epigenomes but their exponential growth in data generation poses significant storage management challenges for research institutions, clinics, hospitals and analysis providers. A typical whole-genome sequencing project generates 100-200 GB of raw FASTQ files, with downstream processing often producing an additional 200-400 GB of intermediate and final analysis files (Stephens et al., 2015). The absence of standardized approaches to sequencing data management results in accumulation of duplicate files, obsolete intermediate outputs, and poorly organized directory structures that impede research productivity and increase infrastructure burden.

Existing solutions focus primarily on general-purpose file deduplication or format-specific compression, but lack the workflow-aware capabilities needed for effective genomics data management. File-level tools typically compute a cryptographic digest (e.g., MD5, SHA-1, SHA-256) and declare files equal when both size and digest match; this is precise, but I/O-bound because the entire file must be read, and weak hashes (MD5/SHA-1) have theoretical collision risks. Fuzzy hashes capture similarity rather than identity and are useful for triage, yet they cannot guarantee that two files are exact duplicates and therefore are unsuitable for safe deletion. In bioinformatics, content-aware canonicalization (e.g., normalizing FASTA records or decompressing GZIP/BGZF to compare payloads) can detect order or container differences, but such normalization changes the equivalence class and may conflict with reproducibility policies that require bit-for-bit provenance. Finally, container/header checks are essential because filenames and extensions are unreliable; header-based typing prevents mislabeling from corrupting dedup or deletion decisions.

To address these limitations, we developed SeqManager, an integrated web-based solution that combines intelligent file classification, optimized duplicate detection, and intermediate file detection specifically designed for sequencing data workflows.

## 2 Methods

### 2.1 System Architecture

SeqManager is implemented as a client-server application using Python Flask for the backend API and React with TypeScript for the frontend interface. The modular architecture consists of four main components: (1) a recursive file scanning engine, (2) an intelligent classification system for sequencing file types, (3) an optimized duplicate detection algorithm, and (4) a tool for detecting intermediate files that can be removed safely.

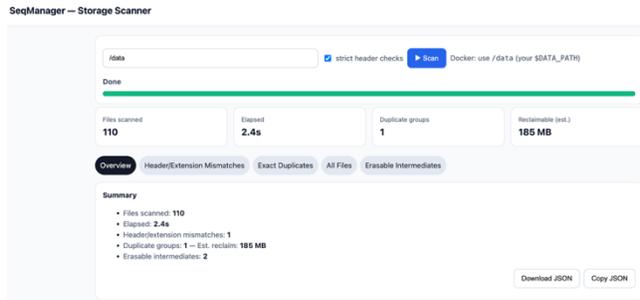

**Figure 1:** Screenshot of the web interface, to scan folders of data and analyse sequencing-related files

### 2.2 File Classification Algorithm

We employ a two-stage classifier:
-path name matching: here the classification engine recognizes major categories of sequencing workflow files based on extension patterns and naming conventions.
-header sniffer: SeqManager verifies the true type using fixed byte counts and offsets. For example, for a VCF file SeqManager reads the first 256 bytes in search of "##fileformat=VCF". This is done for GZIP/BGZF containers, BAM, CRAM, BCF, VCF, SAM, FASTQ and FASTA files. When the header and the extension disagree, the header prevails. Such files are marked with a "header/extension mismatch" flag.

### 2.3 Optimized Duplicate Detection

To search for duplicate files, SeqManager uses a three-tier pipeline.
Files are first bucketed by exact byte size (unsigned 64-bit). Files with different sizes cannot be duplicates and are not compared further.
For each file of size > 196KB we compute a sampled fingerprint by hashing three 64-KB windows (first, middle, last). Each window is hashed with MD5; the three 16-byte digests are concatenated and re-hashed with MD5 to yield a 48-byte to 16-byte fingerprint. Files smaller than 196KB fall back to a single-window MD5 over the entire content. These fingerprints are used only to cluster candidates; they do not justify deletion.
We then perform an exact match by full-file cryptographic hash. All files within a candidate cluster are verified by streaming SHA-256 computed in 4 MiB read blocks. Only files with identical 32-byte SHA-256 digests are labeled Exact duplicates and considered for cleanup.

### 2.4 Erasable Intermediates files

The "Erasable Intermediates" analysis identifies files you can safely remove because an equivalent or superior artifact exists from which they can be deterministically regenerated. It first classifies each file by header bytes (not just suffixes) and normalizes multi-part extensions (e.g., FASTQ.GZ), then groups files by sample stem to build a per-sample inventory. Using conservative rules, it flags SAM when a BAM/CRAM is present (SAM can be re-emitted), uncompressed BAM when a CRAM is retained (BAM is reconstructable using the reference), redundant SRA vs FASTQ(.gz) layers, and trimmed FASTQ when a raw FASTQ plus a manifest of trimming parameters exist. For every candidate it reports the rationale, stated fidelity (bit-or content-equivalent), explicit dependencies (e.g., CRAM and <ref.fa>), and a ready-to-run regeneration command; it never deletes anything automatically, serving as a clear, auditable guide to reclaim storage without risking data loss.

### 2.5 User Interface and Visualization

The web interface provides four views: (1) an overview on scan time, duplicates found and estimated space to be gained, (2) a header/extension mismatch view for files for which the extension does not match the file header type, (3) a view of duplicates with their full puth and the possibility to download these in csv format to delete them, and (4) a list of intermediate files that can be erased.

## 3 Results

We analysed three real use cases that were representative of environments where sequencing data is storedin Table 1. These environments have raw sequencing files across multiple projects with intermediate files but also unrelated files such as pdf figures, documents for billing or scripts.

| Architecture CPU/RAM | Number of raw sequencing files | Number / Total Size(GB) of files to scan | Number of duplicate files | Scan Time (seconds) | Memory Usage (MB) | Space saved (GB) |
|---|---|---|---|---|---|---|
| M1/8GB | 56 | 1670 / 320 | 34 | <1 | 62 | 48 |
| M1/8GB | 80 | 1906 / 1600 | 635 | 2.2 | 81 | 982 |
| 16 c Ryzen 9 7950X / 128 GB | 347 | 34978 / 4154 | 437 | 44.8 | 97 | 822 |
| 48 × (2 × E5-2683 v4 @ 2.1 GHz, 32 c/node) | 347 | 34978 / 4154 | 437 | 37.2 | 112 | 822 |

**Table 1.** Scalability on commonly used Bioinformatics architectures

## Conflict of interest
None declared.

## Data availability
The data underlying this article are available in *https://github.com/AIGeneRegulation/Sequencing-Data-Manager*.